\documentclass{article}
\usepackage{ismir,amsmath,cite}
\usepackage[hyphens]{url}
\usepackage{graphicx}
\usepackage{color}
\usepackage{mathtools}
\usepackage{amsfonts}
\usepackage{enumitem}
\usepackage{nccmath}
\usepackage{booktabs}
 \usepackage{graphicx}
\usepackage{subcaption}

\newcommand\EatDot[1]{}

\DeclarePairedDelimiterX{\infdivx}[2]{(}{)}{%
  #1\;\delimsize\|\;#2%
}

\title{Behavior-based evaluation of session satisfaction}

\multauthor
{Chao Liu \hspace{3.15cm} Zhenzhen Zheng \hspace{3.0cm} Jinkang Jia}
{\bfseries
 Baidu Search Company\\
{\tt\small \{liuchao24, zhengzhenzhen, jiajinkang\}@baidu.com}
}
\def\authorname{Chao Liu, Zhenzhen Zheng, Jinkang Jia}

\begin{document}

\maketitle
\begin{abstract}
Nowadays, web search becomes more and more popular all over the world. Many researchers and developers have done lots of studies on behaviors of search users. In practice, the full understanding of these behaviors can not only help to evaluate the usefulness of newly-developed ranking algorithms and other changes of search engine, but also to guide the growth direction of search engine. As far as we know, most of past work are mainly focused on single search evaluation, which do promote the rapid development of search engine in early stage. However,these page-level behaviors are so limited that can no longer give explicit feedbacks on minor changes of the search engine. We think that it will be more accurate and sensitive when more information on search session are provided. In this paper, a session level evaluation method is proposed. The session-level features are retrieved and carefully analyzed. Some linear and non-linear features which can reflect the final degree of satisfaction are chosen and adopted in evaluation models. A two-layer hybrid evaluation model with different granularity, which can achieve good precision and recall, is designed and trained. Lots of real experiments are evaluated by the model, the result shows it achieved a higher accuracy performance than traditional page-level evaluation metrics. Furthermore, for practical application, it is important to interpret the reason of each session's satisfaction judgement. In all, a session-level evaluation model with improved performance and well capability on interpretation is proposed and applied in real practice in search engine companies. 
\end{abstract}

\section{Introduction}
\label{sec:introduction}

Nowadays, more and more people around the world are used to acquiring information and knowledge through search engine. Users will refer to the search engine naturally when they need to find certain websites, or when they have some informational needs, or when they have some transactions to fulfill, and many other scenes. Search engine companies, such as Google\cite{google}, Baidu\cite{baidu}, etc, receive billions of requests from everywhere on the planet. Moreover, with the rapid rise of the intelligent devices users can express their demands more naturally by voices and images instead of keyboards and mouses traditionally. The sources of the traffic become more various and the companies can make more profits than before. However, these changes also bring many new challenges to the infrastructures and algorithms of the search engine. For example, instead of being satisfied by top several results on PCs or cellphones, users need to be satisfied by only one best result which is accurate and well-organized, in order to be displayed on smaller screens of watches or be read loudly by the speakers. In addition, the interactive behaviors become more complex, and some commands need to be satisfied by multi-round interactions and the context of these questions need to be considered when designing new algorithms.

On the other hand, how to evaluate the search engine comprehensively and accurately under these new changes and trends, is becoming much more important than before. Manual offline evaluation and online experiments are two mostly used methods in traditional web search. Many data annotators are employed and trained to evaluate the search results' qualities by giving the keywords (or queries) and the display results. There are many different evaluation methods and standards, for example, the query-url pair or the side-by-side top-n results evaluation. In practice, there are many other techniques which are used to improve the accuracy and efficiency of the evaluation, such as fitting among multiple scorers, or dynamically adding scorers, etc. We will not discuss these problems in detail here. 

A more comprehensive and low-cost evaluation method is online experiment. Usually two experiments' methods are widely adopted in search environments: A/B test and interleaving\cite{joachims2003evaluating}. A/B test can adapt to many kinds of search engine optimizations, such as display color and pattern changes, highlighting strategies, and sharp ranking adjustments, etc. Interleaving is widely-used in careful refinements of ranking in search result page(SERP) and it is much more sensitive to the minor ranking changes than A/B test. After enough user behaviors/feedback information collected, some statistical metrics, which have been well designed and proved beforehand are calculated. The engineers will make decisions on whether users are in favor of the new changes or dislike them. Therefore, the well-tuning of these metrics is very important and directly influences the correct evolution of search engines.

As far as we know, commonly used metrics in search engines, for instance, click ratio or dwell time in landing pages, are all page-level ones. Generally speaking, the violent increases and decreases of these metrics can help to judge the usefulness of the new algorithms. However, in recent years, we find they become useless gradually with the rapid progress of the search engine. The metrics cannot give directional signals any more on new iterations of search engine. We need to find some new evaluation metrics to accommodate the changes in outer market environment and inner the search engine itself.

The above difficulties motivate us to develop new search engine evaluation methods. Basically, the goal of users is to get what they want satisfactorily and efficiently. That is, each user will perform a group of actions on input, click, browsing, query-change, etc. We call the sequences of these consequent actions a session. We think that it will be more comprehensive and precise to incur more information on successive activities on session instead of only on one page's behaviors as before. The paper will address the full evaluation and application of the session-level metrics.

There are three main contributions of the paper:

1. The establishment of infrastructure for modeling.

In the paper, full procedures and infrastructures on session evaluation for data annotation, feature analysis and selection, modeling and evaluation are proposed. With the in-depth understanding of search users' behaviors, new ideas or features, can be easily added into present model and evaluated quickly. Continuous improvement can be achieved with the deployment of the infrastructure. 

2. Model interpretation. 

Besides the performance of the model, the interpretation capability is of vital importance when put the model into real use. A tool for automatic explanation retrieval is proposed. The experiences on behavior understanding can be accumulated continuously through the tool.
With the accumulated intelligence of search experts, the tool will give more accurate explanation on sessions.

3. Real experiment evaluation and application.

The model and the corresponding explanations can be directed adopted in real world, and it has been proved the session-level evaluation is much better than page-level evaluation by real experiments.. We hope the details caught by session model can further help search engine making progress.

The paper is organized as follows. Some related work will be described in section 2. In section 3 the data source and annotation methods are introduced. The analysis of features and the hints we get from users' behaviors will be discussed in section 4. The construction and selection of the model, and the real-world experiments' evaluation and interpretation on sessions, will be discussed in section 5 and 6 respectively. In section 7, we will conclude our work and discuss some future work.

\section{Related Work}
From the perspective of real applications, we think there are two main concerns which decide the usefulness of the search engine evaluation methods in industry. One is how to accurately evaluate the search engine based on users' consecutive behaviors, and the other is how to interpret the evaluation results of search or session. The latter is also of vital importance because developers need to find out the abnormal behaviors of low satisfaction and try to iterate the algorithms to further optimize user experience.

In this section, we will mainly review related work on behavior-based automatic satisfaction evaluation and the explanatory abilities of these satisfaction models.

\subsection{Satisfaction Evaluation}
Satisfaction evaluation is inherent coming with search engine. "Satisfaction can be understood as the fulfillment of a specified desire or goal" \cite{Kelly2009Methods}. Search engine evaluation can help not only to disclose whether users are satisfied with search results but also to guide the evolution of algorithms in right directions.

Some traditional evaluation methods are mainly based on single search or page view, and this can only reflect the page-level satisfaction to some extent. These methods help a lot in early stage of search engine. The behaviors including clicks, cursor movements, dwell time of landing pages, etc, are the main signals to decide if users are satisfied or not. Click model\cite{Chuklin2015Click} may be one of the first widely-studied models. Probability is usually involved to predict the user satisfaction\cite{cohen1960coefficient}. In order to improve the accuracy, Huang et al \cite{Huang2012Improving} adopt mouse cursor activity information and propose an extended click model. Furthermore, M Ageev et al.\cite{Ageev2013Improving} uses the mouse cursor movements and scrolling over the result document to construct the model which performs better than traditional click models. Besides, there are some other incorporating methods, such as CAS model\cite{Chuklin2016Incorporating}.

However, the particular goal of a user becomes more complex and is usually not fulfilled by only single search. In addition, the page-level metrics become more trivial with the rapid development of search engines. These changes motivate researchers to evaluate search engines on a broader view. One branch is based on click sequences. Sequential pattern method was first introduced by Agrawal et al\cite{Agrawal1995Mining}. On one hand, some probabilistic methods, especially Markov models \cite{Gwadera2005Markov}, are proposed. Hassan et al\cite{Hassan2010Beyond} build another sequence model incorporating time distributions, and their experiments show that sequence-based behavioral models are more accurate than static behavior-based or document-relevance based ones. On the other hand, some methods based on deep learning techniques are adopted in sequential models. Mehrotra et al \cite{Mehrotra2017Deep} use the CNN-LSTM model to construct a deep sequential model and find it performs better than traditional click models and generative probabilistic models\cite{Hassan2012A}.

Although sequential models can achieve better performance than single page evaluation ones, the features used to judge the satisfaction is still limited and it's hard to interpret cases with low satisfaction based only on sequences. On goal level, much more features which may help to describe users' perception on search engines can further be adopted. Jiang et al\cite{jiang2015understanding} use the economic method to reflect the gain and cost of one user and construct a weighted model to evaluate search engine on a graded level. Our work belongs to this category and a comprehensive analysis is conducted on features of China's biggest search engine, Baidu. Also, much attention is put on the model adjustments and interpretations in order to increase the application level in real experiments.

\section{Model Interpretation}
\label{sec:model_interpretation}

\begin{table}
  \caption{Annotated Session Label Distribution}
  \label{table1_session_label_dist}
  \begin{tabular}{ccc}
    \toprule
    Session Label & Satisfaction Level & Proportion\\
    \midrule
    0 & Low & $18.2\%$ \\
    1 & Medium & $18.2\%$ \\
    2 & High & $41.3\%$ \\
    3 & Very High & $22.3\%$ \\
  \bottomrule
\end{tabular}
\end{table}

\section{Data Preparation}

\subsection{Basic Data Description}
A search task, which usually consists one or multiple searches, is called \emph{a session} or \emph{a goal} in the paper. A user may emit one or several tasks each day. Some natural language processing techniques are adopted to split all searches into sessions. Our target is to automatically evaluate if the user is satisfied with the session based on searches and corresponding behaviors.

Search data on page displays and user behaviors are recorded in search logs of Baidu search engine. Some key entries of the logs are: 1) a unique identifier called \emph{goal-id} that denotes which searches belong to the same goal. 2) all results and display style of one search page (SERP). 3) queries and query input types which include manual input, suggestion, related search, history and so on. 4) action types and corresponding timestamps for every behavior in SERP and landing pages. For example, page numbers, urls, ranking positions and dwell time of the results clicked are recorded in order to reappear the scene of the particular user. To make sure the model's robustness on dates, we sampled the data from September 2017 to December 2017.

\subsection{Data Annotation}
To find out the ground truth if users are satisfied with their own search sessions, the most accurate and direct method is to get explicit feedbacks from themselves. For the search requests come from all over the world, it is nearly impossible to do so. Then we try to evaluate the session satisfaction level by multiple annotators. The average evaluation scores of these annotators are regarded as the ground truth.

In order to catch differences among various satisfaction degree and make annotate as simple as possible, four levels are adopted. The label score ranges from 0 to 3, where 0 (Low/L) represents users are dissatisfied with the results and no useful information can be retrieved, and 1(Medium/M) represents users tend to be dissatisfied but some useful information are provided. The other two levels denote users are satisfied with the session generally, and 2(High/H) represents users are basically satisfied but the process is not very smooth or the cost is a little high, and 3(Very High/VH) shows users are very satisfied with very little efforts. In order to analyze more comprehensively, search-level evaluation is also annotated. Three kinds of scores are labeled: 2(very useful), 1(somewhat useful), 0(not useful). Thus, each session comes up with two labels from annotators: session label and search label. In order to reduce the influence of annotators' subjective biases, each session will be evaluated by three annotators independently. We use the average session score of three annotators as the standard session score(referred as \emph{s}), so as to search score per query(referred as \emph{q}).

In practice, annotators will reappear the whole search process of the real users. Each SERP will be shown firstly to help annotator judge if there are some useful information in title or abstract as many searches can be satisfied directly by abstract with no need to click. Then the annotator will click each URL clicked by real user and browse the landing page as real user does. To experience what users get from the whole process, each annotator will be careful enough to get the close evaluation with real users.  

In reality, it is found that there are nearly 32.5\% sessions with only one search, and 67.5\% with multiple searches in Baidu. We will focus more on sessions with more than one search in following sections. Evaluation on single-query sessions will be described in section 5.2 in detail. Finally there are 481 sessions with multiple searches and 120 single-query sessions which are annotated.

Finally, annotators' average score are further discretized into levels. Session label are assigned as follows: 0(\emph{s} $\le 0.67$, Low, 18.2\%), 1($0.67$ < \emph{s} $\le 1.67$, Medium, 18.2\%), 2($1.67$ < \emph{s} $\le 2.67$, High, 41.3\%), 3($2.67$ < \emph{s} $\le 3$, Very High, 22.3\%); similarly, search label are also discretized into 3 grades. Table \ref{table1_session_label_dist} shows the proportion of each session label of dataset.

\begin{table*}
\setlength{\belowcaptionskip}{5pt}
  \begin{subtable}[t]{0.45\linewidth}
  \caption{Search Outcome Features}
  \label{outcome correlation}
    \begin{tabular}{lcccc}
      \toprule
      Features &All&L/M&M/H&H/VH\\
      \midrule
      $Q\_SumClickDwell$ &0.375 & -&0.304& -\\
      $S\_SumClickDwell$ & 0.299 & 0.304 &0.217 & - \\
      $S\_ClickDwell$ & 0.286 & 0.176& 0.212 & - \\
      $QueryInterval$ & 0.152 & - & 0.172 & -\\
      $S\_SumQueryInterval$ & -& -& -& -\\
      $SessionDuration$ & 0.084& 0.075& - & -\\
      $Q\_\#Click \,T \ge40 $& 0.459 & 0.254& 0.394& -0.04\\
      $Q\_\#Click \,T \ge60 $& 0.464 & 0.255& 0.392& 0.003\\
      $Q\_\#Click \,T <15 $& 0.092&0.279 & - & -\\
      $Q\_\#Click \,T <5 $& 0.083& 0.244 & - & -\\
      $S\_\#Click$ & 0.146 & 0.367  & - & -0.23\\
      $S\_\#Click \,T \ge185 $&  0.219 & 0.138 & 0.187 & -\\    
      $S\_\#Click \,T <10$ & - & 0.228 & -& -\\  
    \bottomrule
    \end{tabular}
  \end{subtable}\hfill
  \begin{subtable}[t]{0.45\linewidth}
   \caption{Cost Features}
   \label{cost}
    \begin{tabular}{lcccc}
      \toprule
      Features &All&L/M&M/H&H/VH\\
      \midrule
      $S\_Qlength$ & - & - & - & - \\
      $S\_\#Query$ & -0.119  & 0.235&  -0.246  &  -0.08\\
      $S\_\#InpQuery$ & - & -&  -  & - \\
      $S\_\#HisQuery$ &  - &  0.306  & - & -0.125 \\
      $S\_\#SugQuery$ & - & - & -& - \\
      $S\_\#RSQuery$ &  -0.032 & - &  -0.279  &  0.021\\   
      $S\_AvgClickPos$ &   -0.235  & -&   -0.296  & - \\     
      $S\_MinClickPos$ &  -0.147  & - &   -0.133  & -\\         
      $Q\_MinClickPos$ &   -0.26  & - &   -0.328  & - \\ 
      $Q\_AvgClickPos$ &   -0.25  & - &   -0.309  & - \\ 
      $S\_\#Click$ &  0.146 & 0.367  & - &  -0.23 \\
    \bottomrule
  \end{tabular}
  \end{subtable}

    \begin{subtable}[t]{0.45\linewidth}
      \caption{Changes on Session Features}
      \label{changes correlation}
      \begin{tabular}{lcccc}
      \toprule
       Features &All&L/M&M/H&H/VH\\
      \midrule
      $ \Delta Q\_SumClickDwell$ & - & -& -& -\\
      $ \Delta Q\_\#Click \,T \ge60$ & 0.08 & 0.138 & 0.071 & -0.142\\
      $ \Delta Q\_\#Click \,T <50$ & -& -0.197& 0.15&  -\\
      $ \Delta QEditDistance$ &- & - & -& - \\    
      $ \Delta QJaccardSim$ & - & - & - & - \\  
      $ \Delta Qlength$ &  -0.05 & - & - & -0.132 \\
      $ \Delta QMaxClickPos$ & 0.194 & 0.095 &  0.172 & -\\    
      \bottomrule
    \end{tabular}
  \end{subtable}\hfill
  \begin{subtable}[t]{0.45\linewidth}
  \caption{User Effort Features}
  \label{effort correlation}
   \begin{tabular}{lcccc}
    \toprule
    Features &All&L/M&M/H&H/VH\\
    \midrule
    $S\_\#Query \,NoClick$ & - & - & -0.31  & -0.267\\
    $S\_MaxClickPos$ & -0.198 & 0.268 & -0.275 & -0.145 \\ 
    $Q\_MaxClickPos$ & -0.211  &  0.254 &  -0.256 &  -0.144 \\
    $S\_\#ForwQuery$ & 0.073&  0.148 & - & 0.214\\     
    $S\_MaxQlength$ & -0.198 & 0.123 & -& -0.241\\
    $QEditDistance$& - & -0.282 & -& -0.31\\
    $QJaccardSim$& - & -0.278& - & 0.19\\
    $Q\_\#Click$ &  0.342 & 0.37& 0.235 &  -\\ 
  \bottomrule
  \end{tabular}
\end{subtable}
  \caption{Correlations between features and satisfaction levels. Statistically non-significant values are omitted ("-")}
  \label{tbl:main}

\end{table*}


\section{Features analysis}


When the session data is annotated, it will be used as "ground truth" of the following evaluation model. Various features from users' behaviors are retrieved and the usefulness of them are verified. In the section, some work on feature selection and selection are introduced.

\subsection{Features Selection}

Inspired by \cite{jiang2015understanding}, four aspects of features on sessions are collected to construct the evaluation model.

\textbf{Search Outcome}. Search outcome refers to the degree on which users are truly satisfied. For instance, generally speaking, one user will dwell on the landing page for longer time if more useful information is found on that page. Thus, features on dwell time are devised and retrieved from users' behavior logs.

\textbf{Search Cost}. Search cost is the cost which users pay on the process of satisfaction, e.g., cost of checking unrelated results, cost of changing queries, etc, are considered.

\textbf{User Efforts}. User efforts refers to how much efforts users are willing to make to get satisfied answers. Depending on users' personalities and eagerness of particular goal, these features are somewhat subjective and vary a lot. e.g., the willingness of page forward, the effort on reformatting queries, etc, are considered and retrieved.

\textbf{Changes on Session}. Besides static descriptions on sessions, some dynamic features which reflect the user behavior change, should also be considered. These trend features can provide additional information. In this paper, differences between first and last query search are paid attention to and corresponding features are retrieved. e.g., the change of the number of clicks with long dwell time, the change of query length, etc, are considered.

Due to limited space, complete candidate features are displayed in the appendix in the last page of the paper.

\subsection{Feature Correlation Analysis}

In order to better reveal the relationships between features and the session satisfaction, some preprocessing work should be performed firstly. Some unsupervised outliers detection methods ~\cite{breunig2000lof, rousseeuw1999fast, liu2012isolation} are adopted, as the existence of these extreme values will influence the correlation analysis and the abstraction of latent knowledge, so the recognized outliers are deleted. In addition, some missing value completion techniques and data cleaning are also used for later analysis.

The correlation analysis methods are used to filter the unnecessary features out, and only the features with somewhat obvious relationships with users' satisfaction are retained. The correlation coefficients are calculated not only on the overall datasets but also on groups between the adjacent satisfied levels. Thus, some non-linear features can be clearly shown and adopted. The correlation coefficient and significance test results on four kinds of features are shown from table \ref{outcome correlation} to table \ref{effort correlation} respectively.

\subsection{Summary of Findings}

After careful review of these features and their correlations with satisfaction levels, two main findings are revealed:

1). As shown in table \ref{outcome correlation}, almost all of search outcome features are monotonic linearly related to session satisfaction levels, which means users tend to be more satisfied with the increase of the outcome. 
Take $Q\_\#ClickT>=60$ as example, it isn't difficult to understand that users are more satisfied when these kind of "long" clicks are more in the session. In particular, this feature is much more indicative in L/M and M/H levels compared to H/VH. That is, different features act very differently in classifying different satisfaction levels.

\begin{figure}
 \hfill\begin{minipage}{.5\textwidth}\centering
\begin{subfigure}[t]{1.5in}
    \centering
    \includegraphics[width=1.6in,height=1.25in]{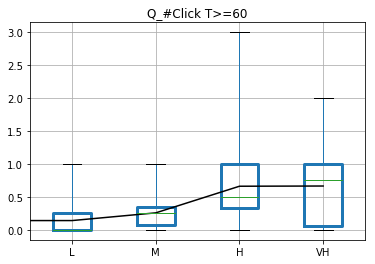}
    \includegraphics[width=1.6in,height=1.25in]{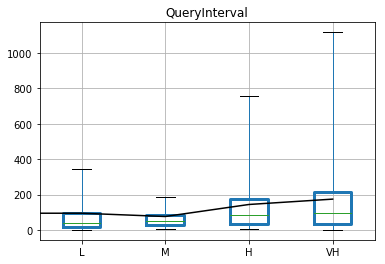}
    \caption{Linear relationship}
\end{subfigure}
\begin{subfigure}[t]{1.5in}
    \centering
    \includegraphics[width=1.6in,height=1.25in]{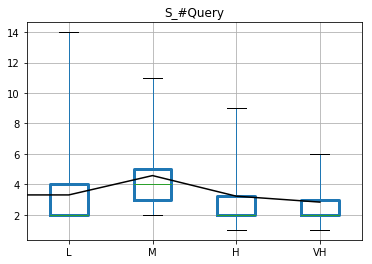}
    \includegraphics[width=1.6in,height=1.25in]{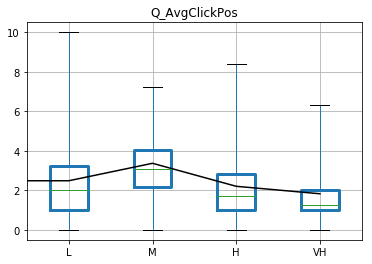}
    \caption{Non-linear relationship}
\end{subfigure}
  \caption{Examples of linear or non-linear features}
  \label{monotonic-relation}
 \end{minipage}

\end{figure}

Most of search cost and user effort features show non-monotonic linear relations with session satisfaction levels. This can be explained as follows. On one hand, when faced with terrible unrelated search results, users will give up quickly; but on the other hand, users will directly find out what they want and get satisfied quickly when faced with well-formatted and relevant results. Both of these two situations will lead to low value in search cost and user effort features. Under these two extreme situations, users will spend more patience and attention to search for what they want, which will lead to higher values. The example features on both linear and non-linear relations are show in Figure \ref{monotonic-relation}.

\begin{figure}
\includegraphics[width=2in,height=2in]{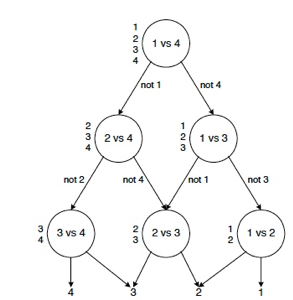}
\caption{Structure of DAG-SVM}
\label{dag-svm}
\end{figure}

2). There are some interesting search behavior differences between Chinese and English speaking users by comparing correlation analysis results in \cite{jiang2015understanding} and our work.

$Q\_\#Click \,T <5$, $Q\_\#Click \,T <15$, $S\_\#Query$ are all negative correlation between Medium and High groups in Bing, on the opposite, Baidu are both positive correlation between Low and Medium groups. Which means that these features are small when satisfaction level is Low, but bigger when satisfaction is Medium in Baidu, similarly these features are higher when satisfaction level is Medium and smaller when satisfaction is High in Bing. It is indicated that English-speaking users are more patient when faced with seemingly unrelated results. Chinese users tend to give up earlier and depend more on the snippets in their judgments. English users are willing to spend more time on each click and reform more queries to search for good results in Bing, as the improvement of search results, these features become smaller, satisfaction level is higher.

Another thing is that in High and Very High groups, the feature $QJaccardSim$ is smaller in Bing. The reason may lie in different language grammer. For English users, to find more satisfied results, English users may need to modify original query thoroughly to synonyms or others. While Chinese users may change a little, such as some prefixes or suffixes, to form a new query with different meaning.

\section{Model Construction}

After deep insights and understanding are obtained on how the behaviors influence the final satisfied levels, models will be built to predict if the new given session is satisfied or not. Because there exists many features which cannot be correctly retrieved in single-query sessions, we divide the model into two parts. One is responsible for complex sessions with multiple searches, and the other is used to process relatively simple one-query sessions. These two parts will be introduced in 5.1 and 5.2, respectively. In 5.3, the combined final model is for real application is proposed.

\subsection{Session Modeling for Multiple Queries}

\begin{table*}
\setlength{\belowcaptionskip}{2pt}
  \begin{subtable}[t]{0.45\linewidth}
  \caption{Basic binary classification models (One vs One)}
  \label{performance of binary}
  \begin{tabular}{cccc}
    \toprule
    Active/Negative class & Precision & Recall& F1-score \\
    \midrule
    2 vs 3 & 0.669 & 0.667 &0.668\\
    1 vs 3 & 0.717& 0.706&0.705\\
    1 vs 2 & 0.753 & 0.758&0.748\\
    0 vs 2 & 0.784&0.778 &0.780\\
    0 vs 3 & 0.779&0.771 &0.772\\
    0 vs 1 & 0.772&0.756 &0.754\\
  \bottomrule
\end{tabular}
\end{subtable}\hfill
\begin{subtable}[t]{0.45\linewidth}
  \caption{Basic binary classification models (One vs Rest)}
  \label{performance of binary one vs rest}
  \begin{tabular}{cccc}
    \toprule
    Active/Negative class & Precision & Recall& F1-score \\
    \midrule
    0 vs Rest & 1.000 & 0.333&0.500\\
    1 vs Rest & 0.333& 0.0769&0.125\\
    2 vs Rest & 0.540& 0.714&0.615\\
    3 vs Rest & 0.750&0.273 &0.400\\
  \bottomrule
\end{tabular}
\end{subtable}
\begin{subtable}[t]{0.45\linewidth}
  \caption{Combined binary models}
  \label{performance of combined binary}
  \begin{tabular}{lccc}
    \toprule
    Combined method &  Precision & Recall  & F1-score\\
    \midrule
    One vs One & 0.587&0.578 &0.556\\
    One vs Rest & 0.621& 0.438& 0.457\\
    DAG-SVM & 0.546& 0.571& 0.553\\
    DAG-SVM(sat vs. dissat) &0.598&0.593&0.594 \\
  \bottomrule
\end{tabular}
\end{subtable}\hfill
\begin{subtable}[t]{0.45\linewidth}
  \caption{Comparision among different multi-class models}
  \label{multi-class models}
  \begin{tabular}{lccc}
    \toprule
    Models& Precision  & Recall & F1-score\\
    \midrule
      Logistic Regression&0.506&0.487&0.465 \\
      LinearSVM&0.465&0.462&0.444 \\     
      Random forest&0.568&0.556&0.537 \\
      Gbdt&0.650&0.600&0.620\\     
      Xgboost&0.678&0.647&0.662 \\ 
  \bottomrule
\end{tabular}
\end{subtable}
  \caption{Comparison of different model performance}
  \label{all performance}
\end{table*}

Classification is a classic problem in machine learning. Binary classification methods, such as logistic regression, support vector machine, etc., and multiple ones, such as decision tree, random forest, etc., are adopted and tested in our modeling work. Finally, a combined model with binary and multiple classifiers are proposed and better performance are achieved. We will discuss these work in detail in following subsections.

\subsubsection{Binary Classifier Models}
\ 

As mentioned above, there exists some important features with good discriminative power which are only statistically significant on limited levels. To make full use of these detailed information, it may be better to construct binary classifier models on levels with different satisfaction. For there are four satisfied classes (labels), prediction results of these basic binary classifiers should then be combined together to achieve final predictions. Three main combination strategies are adopted as described in the following:

\textbf{One vs. One}: Given $N$ labels, there are in total $C_N^2$ basic binary classifier models. In prediction phase, each session will firstly pass through all basic models. Secondly a simple voting algorithm is adopted to produce final prediction of satisfaction level. In table \ref{performance of binary}, the precision and recall of these basic models are denoted. The precision ranges from 0.7 to 0.85, which is higher than the consistency of human annotators. However, the second-phase voting doesn't work well and the performance becomes worse for the errors of each basic model may be magnified during voting.

\textbf{One vs. Rest}: Unlike One vs. One, in training phase One vs. Rest model uses sessions with certain particular label as positive and all the rest as negative, thus $N$ models are built in total. the performance of $N$ models can be seen in table \ref{performance of binary one vs rest}. $N$ prediction results which indicate if the session belongs to the target label or not are produced and the positive label with maximum probability is decided. Due to the imbalance of the training instances, some re-sample techniques should be used to eliminate these biases.
It can be concluded that both the basic model and the final model don't work as well as One vs. One mentioned above. We think there may be more ambiguities incurred in "rest" group and it's more difficult to separate the target label from the other.

\textbf{DAG-SVM}\cite{platt2000large}: This method originates from traditional SVM and can support multiple-label classifications. Unlike the flat voting infrastructure as above two kinds, this method uses the directed acyclic graph which incurs the dependencies upon different basic binary classifiers. The structure of DAG-SVM is like a decision tree as shown in Figure \ref{dag-svm}. The root binary classifier which usually has the best performance is firstly passed, and the prediction result is given. Based on the result label, the tree path is then determined. The drawback of the method is that the error will be accumulated from root to leaves, which will deteriorate the overall performance of the model.

In addition, another DAG-SVM structure (sat vs. dissat) is also tested.  Session instances of L/M and H/VH are grouped because the former two kinds of sessions tend to be satisfactory while the latter be dissatisfactory. The root binary classifier is used to classify a new session into satisfied or not, and the rest works as before. As expected, it achieve better precision and recall than the ordinary DAG-SVM. This motivates us that the performance can be improved by adjusting the structure of models which may be more suitable for classification tasks.

In table \ref{performance of combined binary}, it can be drawn that all direct combinations of basic binary classifiers perform not well enough to put into real use. Some other methods and optimizations should be raised.

\subsubsection{Multiple Classifier Models}



\ 

As described in last subsection, the direct combinations of basic binary classifiers don't work very well. We will try to adopt the classifiers which support multi-label predictions in this part.

Many widely-used classification methods are tested, and the result in table \ref{multi-class models} shows that xgboost achieves best performance in our environment. In general, it's clear that non-linear models which compose random forest, gbdt and xgboost, can achieve better performance than linear ones including logistic regression and linearSVM.

It can be concluded that the overall performance is better than any combination of binary classifiers. However, the precision and recall are not high enough to be applied in real world. After careful analysis on concrete sessions, it's found that the xgboost model works well in distinguishing the satisfied (label 2 and 3) from the unsatisfied (label 0 and 1), but the model is not good at discriminating between the more granular levels, such as L/M and H/VH pairs. This motivates us to combine this model with the binary classifiers mentioned above, and the combinations are expected to achieve better performance if the model can make full use of the information at all granularities.


\subsubsection{Hybrid Model}
\ 

The hybrid model will be introduced in this subsection. As discussed above, a two-layer model is proposed to better adapt the characteristics of our applications. 

\begin{figure}
\includegraphics[width=3.5in,height=1.8in]{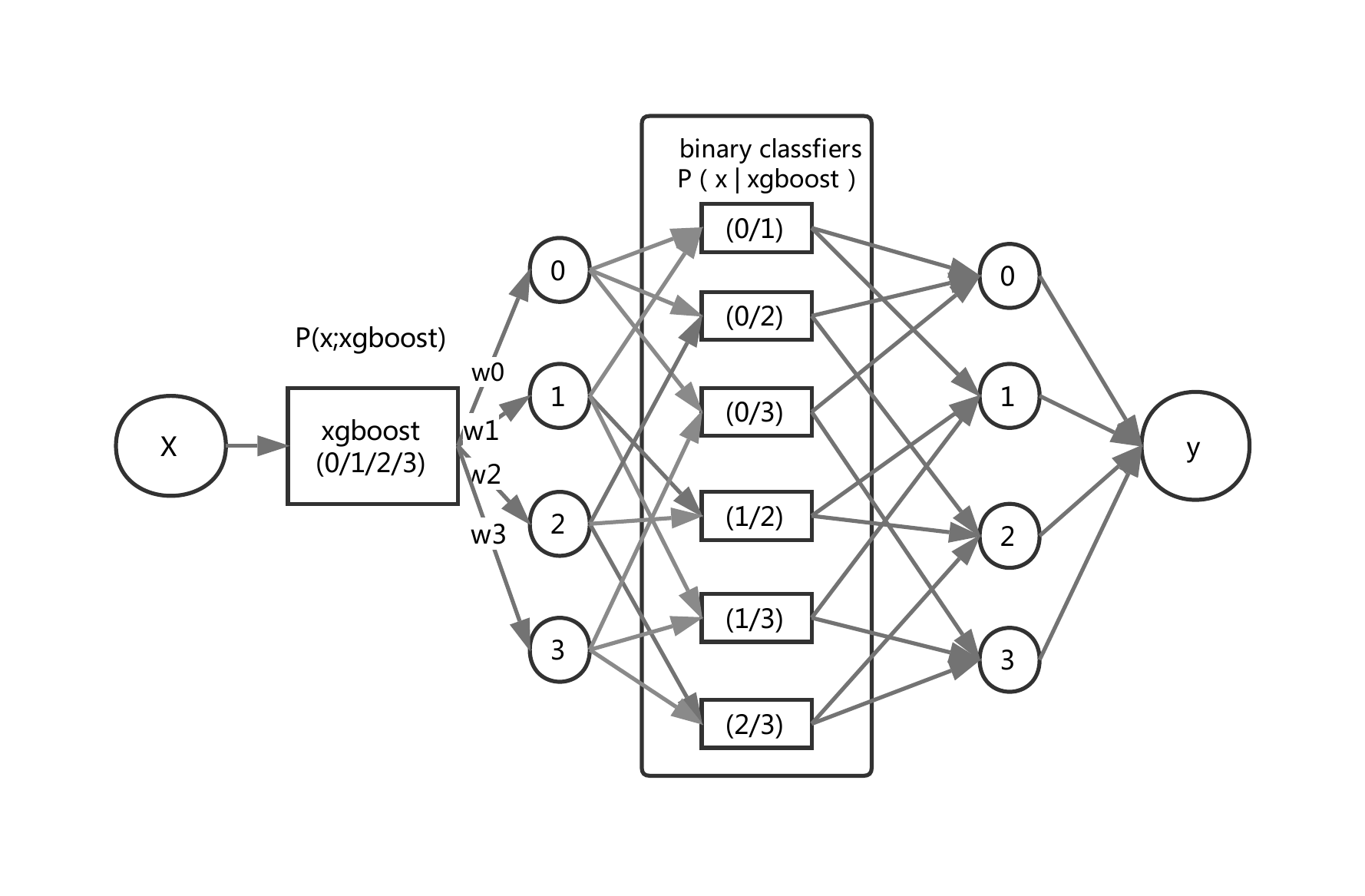}
\caption{Structure for Hybrid Model on sessions }
\label{Hybrid Model}
\end{figure}

In prediction phase, each session sample will firstly go through the multiple classification model (e.g. xgboost) layer to make a general judgment, and probabilities of each satisfaction level are given. Then the conditional binary classifiers are used to make a second round decision. The structure of the model is plotted in Figure \ref{Hybrid Model}. Suppose $P_i(X)$ denotes the probability that session sample X is predicted as label $i$. The formal combined Bayesian-like model is defined as follows.

$$P_i(X)= \sum_{j=0}^nw_iP_j(X;xgboost)P_i(X| j, xgboost) $$
$$Label(X) = \max_{i=0..n}P_i(X)$$

\begin{table}
  \caption{Hybrid model for sessions}
  \label{hybrid model performance}
  \begin{tabular}{cccc}
    \toprule
    Label & Precision  & Recall & F1-score\\
    \midrule
      0&0.917&0.786&0.846 \\
      1&0.731&0.950&0.826 \\
      2&0.632&0.828&0.716 \\
      3&1.00&0.278&0.435 \\ 
      total&0.787&0.728&0.756 \\
  \bottomrule
\end{tabular}
\end{table}

As shown above, $P_j(X;xgboost)$ is the output probability of label $j$ generated by xgboost, and $P_i(X| j, xgboost)$ is the conditional probability predicted by binary classifiers to be label $i$ based on the condition xgboost predicts X to be label $j$. $w_i$ is the weight of the two model combination and can be regarded as the accuracy parameter of the prediction path. $w_i$ is automatically searched in the parameter space so as to achieve the best performance (e.g., F1-score). $Label(x)$ is the final classification label for sample X. It can be seen from Table \ref{hybrid model performance} that the performance of the hybrid model increases a lot compared to binary or multiple classifiers beforehand.

In practice, the model can be adjusted according to the application requirements. For example, the above weight $w_i$ can be optimized due to the different tolerance degree of misclassification. Similarly, for computing convenience, the structure of the model can also be simplified and some paths can be removed. Only the most possible misclassified binary classifiers are remained based on confusion matrix. For example, the L/VH binary classifier needs to be careful examined because sometimes it's difficult to accurately predict the label, for sometimes users can be directly satisfied by abstracts in SERP without any clicks. After these optimizations, the calculation performance can be further improved with almost similar precision and recall.

\subsection{Session Modeling for Single Query}

In single-query modeling, many features, such as those reflecting changes in search process, can no longer be correctly retrieved. For this part only the most correlated features with single-query satisfaction are remained and a simple decision tree model is built to predict the final label of these single-query sessions. The left features are mainly composed of three categories: 1) features on dwell time (e.g., more "long" clicks generally mean users are more easy to get satisfied), 2) clicks (e.g., no click usually means dissatisfied) and 3) queries (e.g., short queries usually are easier to be satisfied comparing to long ones). 

In addition, some techniques, such as recognizing if the requirements can be directly satisfied by snippets or abstracts, are also studied and adopted in real applications. Features related to clicks and consequent behaviors, become useless. Rule-based methods are mined and summarized by analysis. For example, the search frequency of each query is considered, and it is likely the user can be immediately satisfied for there are many specialized optimizations on these hottest queries. Also, many other features, such as the duration of session, the statistical click ratio of the query, are also taken into account. In all, rather high precision (0.88) and recall (0.90) can be achieved for single-query sessions from table \ref{final performance}

\subsection{Final Session Modeling}
Finally, the hybrid model for long sessions and the tree-like model for single-query session are combined together and the final results are shown in table \ref{final performance}. The performance with the precision of 0.82 and the recall of 0.79, is higher than human annotators and can be used in overall evaluation of search engines.

\begin{table}
  \caption{Independent/Hybrid models for all sessions}
  \label{final performance}
  \begin{tabular}{cccc}
    \toprule
    Session type& Precision  & Recall & F1-score\\
    \midrule
     Multiple queries session &0.787&0.728&0.756 \\
     Single-query session&0.885&0.902&0.893 \\
     Total & 0.819& 0.785&0.802\\
  \bottomrule
\end{tabular}
\end{table}

\begin{table*}
  \caption{Example explanations on classified sessions} 
  \label{top rules}
  \begin{tabular}{p{1.5cm}p{8cm}p{6cm}}
    \toprule
    Satisfaction & Behaviors & Description \\
    \midrule
      Very High & short length query, clicks on the front position, small number clicks. & quickly satisfied without cost  \\
      Very High & many clicks, very long dwell time, clicks on the front position. & satisfied with many long clicks \\
      Low & short dwell time, few clicks, many no click queries, similar adjacent queries. & many no-click queries and leave disappointed \\
      Low & low quality results, high cost of assessing, effort changed lower. short-time clicks, big number of clicks, trend to ignore the back ranking results. & many deceptious clicks without real content \\
      High &  many clicks, long dwell time, long duration of session, has the longer trend of dwell time. & increasing search experiences among queries with longer dwell time and clicks.\\
      Medium & few clicks per query, clicks on the back position, few long dwell time clicks. & struggling to find little information on low-rank clicks.\\
  \bottomrule
\end{tabular}
\end{table*}

\section{Experimental Evaluation}

\begin{table}
  \caption{Evaluation results between session and page-level metrics}
  \label{experiments GSB}
  \begin{tabular}{lccc}
    \toprule
    Page-level Metrics & \#Good & \#Same  & \#Bad  \\
    \midrule
     has\_click\_ratio  & 58 & 426 & 10  \\
     click\_ratio   & 60 & 427 & 6  \\ 
     long\_click\_ratio & 100 & 394 &4  \\
  \bottomrule
\end{tabular}
\end{table}

In order to further evaluate the performance of the session model in real world, some real experiments are adopted. Everyday there are many ongoing A/B tests in Baidu. Some of them are randomly chosen and the difference between the treatment group and the control group is calculated for each A/B test. The positive difference value means that the treatment group is very promising and outperforms the control group, while the negative means the treatment group don't work very well as the control group. The traditional page-level metrics, such as has-click ratio, click ratio, and long-click ratio, are calculated as reference metrics, and the average score of our session model is also calculated for comparison.

\subsection{Comparison with Page-level Metrics}

In order to compare metrics at same granularities, the average of page-level metrics are calculated on session level too. Generally speaking, it's shown that some page-level metrics ($has\_click\_ratio$ and $click\_ratio$) and our session metric are weakly correlated, for click information may be the most weighted features no matter in page or session level metrics. 
However, the differences between the experimental and the control group vary a lot in A/B tests. In order to measure which metric is more accurate, about 500 sessions are randomly sampled from experiments. These sessions are evaluated by annotators and labeled as three grades (good, same, bad). Good means that the session model's judgment is better than page-level metrics, and bad means worse. From the distribution in table \ref{experiments GSB}, it can be concluded that our model performs much better than page-level judgments. From the analysis of the good cases, two advantages are found apparently. One is that page-level is single-faceted on simple behaviors such as clicks and dwell times, while session model do make a comprehensive evaluation. The other is that page-level metric only considers the current behavior without the context information.

\subsection{Model Interpretation}
After the model is built, each new-coming session can be evaluated and its' satisfaction judgment is immediately given. However, it is also very important to give the reason why users are satisfied or not in detail so as to help developers to understand what users' behaviors mean. This will help a lot for engineers to make further optimizations. Interpretability of the "black-box" model output should be increased. 

To achieve this goal, three steps are processed in order. Firstly, signal features, which really function for prediction, should be retrieved from user's behaviors of the session. LIME (short for "Local Interpretable Model agnostic Explanation")\cite{ribeiro2016should} is an open-source tool which is used to explain the machine learning "black-box" model's prediction result, is adopted. Secondly, values of these signal features are discretized according to the feature distribution calculated beforehand. Thus, some quantitative explanations on each session, such as "a session with very short queries and ultra-low page-level maximum click position is labeled as satisfaction level of Very High", can be concluded. After this step, the session-specific explanation is given. Thirdly, rules retrieved automatically by machine need to be abstracted further in order to locate problems at higher level. The features are further generalized according to experts' judgment, such as cost of querying, quality of landing pages, patience of target users, and so on. Thus, the top-bottom explanation system can be established and developers can easily trace what problems their new ranking algorithms incur. There are 42 top rules, which cover about 98\% of all sessions, are retrieved from our datasets. Some significant explanations are listed table \ref{top rules}.


\subsection{Case Analysis}

A demo case is shown in table \ref{Session demo case}. The user's goal is to find out the whole process of how to register apple id on cellphone. There are three queries the user emits in total. The first query is "apple" and the user clicks an advertisement of apple's official website, which is promoting new products such as iphone X. This has no relevance with user's demand and the user then move to the search box to emit the second query "apple id registration" by clicking the suggested queries. This time the user watches some video tutorials and clicks the official website. The website link doesn't show the registration entrance but some information on how to find back registered id or password. The user is a little confused and emit the third query "apple id registration tutorial" in a clearer way. This time the user finds some detailed operational procedures on baidu experience related products. At last the user turns to the website to fulfill his or her registration. The whole process is so complex that the 'has click' or 'click ratio' metrics can not reflect if the user is really satisfied. The session model utilizes the dwell time, input methods, and changes among different queries and gives 'High' satisfied result.

Through explanation mechanism, the rule for evaluating is given: increasing search experiences among queries with longer dwell time and clicks. And the description for the rule is: high quality results are finally found. Though struggling at the beginning, qualities of results become better and better during whole process.


\section{Conclusions and Future Work}


In this paper, a comprehensive evaluation mechanism on session-level is proposed in web search, and the procedures and infrastructure are established too. The features are carefully devised and retrieved, and the relationships between these features and the final satisfaction levels are studied. A hybrid two-layer classification model which is devised to make full use of the granular information is proposed based on selected features. The model achieves promising performance and has been put into real use in Baidu. By comparing with old page-level metrics, session-level judgments are more comprehensive and context-aware. In addition, an explanation system for automatic rule retrieval and abstraction is raised. We hope that this work will greatly help search engine make further improvement in the future.

There are some other optimizations which can be well-studied in the future. Firstly, to improve abilities of model's description level, new features can be continuous tested and introduced into the model. For example, users' profiles, industries of queries, and many others may help a lot in predicting the final satisfaction levels. Secondly, as for features' analysis, only one-dimension features are studied and described in the paper, there may exist some linear or non-linear relationship between satisfaction grades and two or even higher dimension combinations. These latent relationships need to be discovered. Thirdly, we think that the explanations on particular experiment can be further aggregated and abstracted. These high-level knowledge will directly show the advantages and disadvantages of the new algorithms, which may result in higher efficiencies for iterations of them.

\appendix


\begin{table*}
  \caption{All retrieved feature list}
  \label{feature list}
  \begin{tabular}{ll}
    \toprule
     Feature & Meaning\\
    \midrule
    \textbf{search outcome} & \\
    $Q\_SumClickDwell$ & the average of sum of click dwell time in a query\\
    $S\_SumClickDwell$ & the sum of click dwell time in a session\\
    $S\_ClickDwell$ & average dwell time of a click\\
    $QueryInterval$ & average query dwell time in a session\\
    $S\_SumQueryInterval$ &the sum of query dwell time\\
    $SessionDuration$ & the duration time of a session\\
    $Q\_\#Click \,T \ge40$ &  the average of clicks in a query that dwell time $\ge 40s$ \\
    $Q\_\#Click \,T \ge60$ &  the average of clicks in a query that dwell time $\ge 60s$\\
    $Q\_\#Click \,T <20$ & the average of clicks in a query that dwell time < $20s$\\
    $Q\_\#Click \,T <5$ & the average of clicks in a query that dwell time < $5s$\\
    $S\_\#Click$ & the clicks in a session \\
    $S\_\#Click \,T \ge185$ & the clicks in a session that dwell time$\ge185s$\\      
    $S\_\#Click \,T <10$ & the clicks in a session that dwell time<$10s$\\  
    \textbf{search cost} & \\
    $S\_Qlength$ & average query length in a session\\
    $S\_\#Query$ &  the number of queries in a session\\
    $S\_\#InpQuery$ & the number of input query in a session\\
    $S\_\#HisQuery$ & the number of history query in a session\\
    $S\_\#SugQuery$ & the number of suggestion query\\
    $S\_\#RSQuery$ & the number of related search query in a session\\   
    $S\_AvgClickPos$ & the average rank position of clicks in a session\\     
    $S\_MinClickPos$ & the minimum rank position of clicks in a session\\         
    $Q\_MinClickPos$ & the average of minimum rank position of clicks in each query\\ 
    $Q\_AvgClickPos$ & the average of average rank position of clicks in each query \\ 
    $S\_\#Click$ & the number of clicks in a session\\ 
    \textbf{user effort} & \\
    $S\_\#Query \,NoClick$ & the number of queries without clicks\\
    $S\_MaxClickPos$ & the maximum rank position of clicks in a session\\ 
    $Q\_MaxClickPos$ & the average of maximum rank position of clicks in each query\\
    $S\_\#ForwQuery$ & the number of page forward \\     
    $S\_MaxQlength$ & the maximum Query length in a session\\
    $QEditDistance$& the average of edit distance of adjacent queries\\
    $QJaccardSim$& the average of Jaccard Similarity of adjacent queries\\
    $Q\_\#Click$ & the average clicks in each query\\     
    \textbf{outcome and effort change} & \\
    $ \Delta Q\_SumClickDwell$ & the difference of the last and first query in the sum of dwell time\\
    $ \Delta Q\_\#Click \,T \ge60$ & the difference of the last and first query in clicks that $\ge 60s$ \\
    $ \Delta Q\_\#Click \,T <50$ &the difference of the last and first query in clicks that <50s\\
    $ \Delta QEditDistance$ & the difference of edit distance of last and first adjacent queries \\
    $ \Delta QJaccardSim$ & the difference of Jaccard similarity of last and first adjacent queries\\  
    $ \Delta Qlength$ & the difference of query length between last and first query\\
    $ \Delta QMaxClickPos$ & the difference of maximum rank position of clicks between last and first query\\  

  \bottomrule
\end{tabular}
\end{table*}



\bibliographystyle{ACM-Reference-Format}
\bibliography{references}

%
%
%
%

\end{document}